\documentclass[journal]{IEEEtran} 

\ifCLASSINFOpdf
\usepackage[pdftex]{graphicx}
\DeclareGraphicsExtensions{.pdf,.jpeg,.png}
\else
\usepackage[dvips]{graphicx}
\fi

\usepackage{epstopdf}
\usepackage[cmex10]{amsmath}
\usepackage{amssymb}
\usepackage{cases}
\usepackage{wasysym}
\usepackage{relsize}
\usepackage{algorithm}
\usepackage{algorithmic}
\usepackage{array}
\usepackage{cite}
\usepackage{color}
\usepackage{url}
\usepackage{bm}
\usepackage{enumerate}
\usepackage[capitalize]{cleveref}
\usepackage{amsfonts}
\usepackage{epsfig,latexsym}
\usepackage{balance}

\usepackage{float} 
\usepackage{subcaption}

\allowdisplaybreaks[4]

\newtheorem{remark}{Remark}

\usepackage{wrapfig}
\newcommand{\figref}[1]{Fig.~\ref{#1}}

\usepackage{subfig}
\DeclareSubrefFormat{parens}{#1(#2)}
\renewcommand{\algorithmicrequire}{\textbf{Input:}}
\renewcommand{\algorithmicensure}{\textbf{Output:}}
\makeatletter

\renewcommand*{\@opargbegintheorem}[3]{\trivlist
      \item[\hskip \labelsep{\bfseries #1\ #2}] \textbf{(#3):}\ }
\makeatother     
\usepackage{fancyhdr}
\begin{document}

\title
{\textcolor{red}{Non-Uniform Activation Aided Distributed Generalized Spatial Modulation for DMIMO Systems}}
\author{Haojin Li, Chen Sun$^{*}$, ~\IEEEmembership{Senior Member, IEEE}, Rang Su, Zhaojian Liu, Wenqi Zhang, and\\ Haijun Zhang,~\IEEEmembership{Fellow, IEEE} 
\thanks{
Haojin Li, and Haijun Zhang are with University of Science and Technology Beijing, Beijing, China, 100083 (E-mail: zhanghaijun@ustb.edu.cn.). $^{*}$ Corresponding author.

Haojin Li, Chen Sun, Wenqi Zhang are with Sony China Research Laboratory, Beijing, 100027, China (E-mail:\{haojin.li, chen.sun, wenqi.zhang\}@sony.com). 

Rang Su is with the Beijing University of Posts and Telecommunications, Beijing, China (e-mail: surang@bupt.edu.cn).

Zhaojian Liu is with the Beijing Jiaotong University, Beijing, China (e-mail: 23120090@bjtu.edu.cn).

}
}
   
\maketitle
\thispagestyle{fancy}
\fancyhead{}
\chead{}
\rhead{}
\lfoot{}
\cfoot{\thepage}   
\rfoot{}
\fancyhead[L]{Submitted to IEEE Wireless Communications Letters}
\renewcommand{\headrulewidth}{0pt}
\newpage

\begin{abstract}
	\color{red}This letter investigates a non-uniform activation aided distributed generalized spatial modulation (DGSM) scheme in a downlink distributed multiple-input multiple-output (DMIMO) system formed by geographically separated transmission reception points (TRPs). Unlike conventional uniform-activation DGSM, the proposed scheme assigns adaptive activation probabilities to different TRP subsets, thereby jointly exploiting the data-domain transmission capability and the index-domain information carried by cooperative subset selection. We derive a mutual-information upper bound under non-uniform activation and formulate a joint activation-probability and power allocation problem subject to a long-term average power constraint. To solve this problem, we develop a semi-closed-form joint probability and power optimization (SC-JPPO) algorithm, where the optimal activation probabilities are shown to follow a Softmax function of the effective utilities. Simulation results demonstrate that the proposed scheme improves spectral efficiency under the same activation scale and power budget compared with existing benchmarks.\normalcolor
\end{abstract}


\section{Introduction}
{\color{red}{Distributed} multiple-input multiple-output (DMIMO) systems employ geographically separated transmission points to form a virtual antenna array and thereby enhance spectral efficiency, coverage, and spatial diversity \cite{Lee2012CoMPDeploy,Lee2012CoMPSystems,Sun2013CoMP}. A distinctive feature of DMIMO is that the active transmission points can be selected from a distributed set according to the channel condition, the activation scale, and the available transmission resources. This creates an opportunity to exploit the active-node pattern itself as an information-bearing resource, in addition to the conventional data symbols transmitted over the selected distributed antennas. However, most existing cooperative transmission designs still treat the active-node pattern primarily as a scheduling or beamforming variable, rather than optimizing it as part of the information-bearing signal structure.}

Spatial modulation (SM) provides a different design philosophy by embedding
additional information into the indices of transmit antennas \cite{Mesleh2008SM,
	DiRenzo2011SMSurvey}. This index-modulation principle enables an attractive
tradeoff among spectral efficiency, energy efficiency, and implementation
complexity, and has been extensively investigated in emerging wireless
systems \cite{Wen2019SMSurvey}. To further enhance the index-domain
transmission capability, generalized spatial modulation (GSM) activates
multiple transmit antennas simultaneously and conveys information through
antenna-combination indices \cite{Younis2010GSM,Wang2012MASM}. GSM has also
been extended to large-scale multiuser multiple-input multiple-output (MIMO) systems, where the enlarged
spatial-index set offers additional spectral efficiency benefits
\cite{Narasimhan2015GSM}. {\color{red}Recent RIS-aided index-modulation and spatial-modulation schemes have further
exploited reconfigurable propagation environments to improve spectral
efficiency and reliability, including received adaptive SM designs \cite{10978803,11506249}.}

The concept of index-domain transmission has also been introduced into
cooperative networks. Distributed spatial modulation (DSM) exploits the
indices of distributed relays to convey information and achieve cooperative
diversity \cite{Narayanan2016DSM}. More recently, DGSM has been investigated
for relay networks with multiple simultaneous relays, extending GSM to
distributed cooperative nodes \cite{Elganimi2020DGSM}.
Nevertheless, most existing DGSM-related schemes assume uniform activation
over all admissible node subsets. Such a uniform design maximizes the
selection entropy but ignores the heterogeneous channel qualities of
different cooperative node combinations.

{\color{red}This letter proposes a non-uniform activation aided DGSM scheme in
the considered DMIMO setting. Unlike conventional uniform-activation DGSM,
where each admissible node subset is assigned an equal activation probability,
the proposed scheme adaptively optimizes the subset activation probabilities
according to the channel conditions, thereby providing a flexible
tradeoff between index-domain transmission and channel-quality-oriented node
selection. Based on the proposed model, we derive a mutual-information upper
bound and formulate a joint activation-probability and power allocation
problem under a long-term average power constraint. To solve this problem, we
develop a semi-closed-form joint probability and power optimization (SC-JPPO)
algorithm, which yields a global water-filling structure for power allocation
and a Softmax structure for the optimal activation probabilities.}

\textbf{\textit{Notations}}: $[\cdot]^\text{T}$ and $[\cdot]^\text{H}$ denote the transpose and conjugate transpose, respectively. $s$, $\boldsymbol{s}$, and $\boldsymbol{S}$ denote a scalar, vector, and matrix, respectively. $\|\boldsymbol{s}\|_2$ denotes the $l_2$ norm. $\mathrm{Tr}(\cdot)$, $\mathrm{diag}(\cdot)$, and $\det(\cdot)$ denote the trace, diagonalization, and determinant operators, respectively. $\boldsymbol{I}_N$ represents the $N\times N$ identity matrix. $\left(a\atop b\right)$ represents the number of combinations of selecting \(b\) elements from a set of \(a\) elements.

\begin{figure}[!t]
	\centering
	\includegraphics[width=0.9\linewidth]{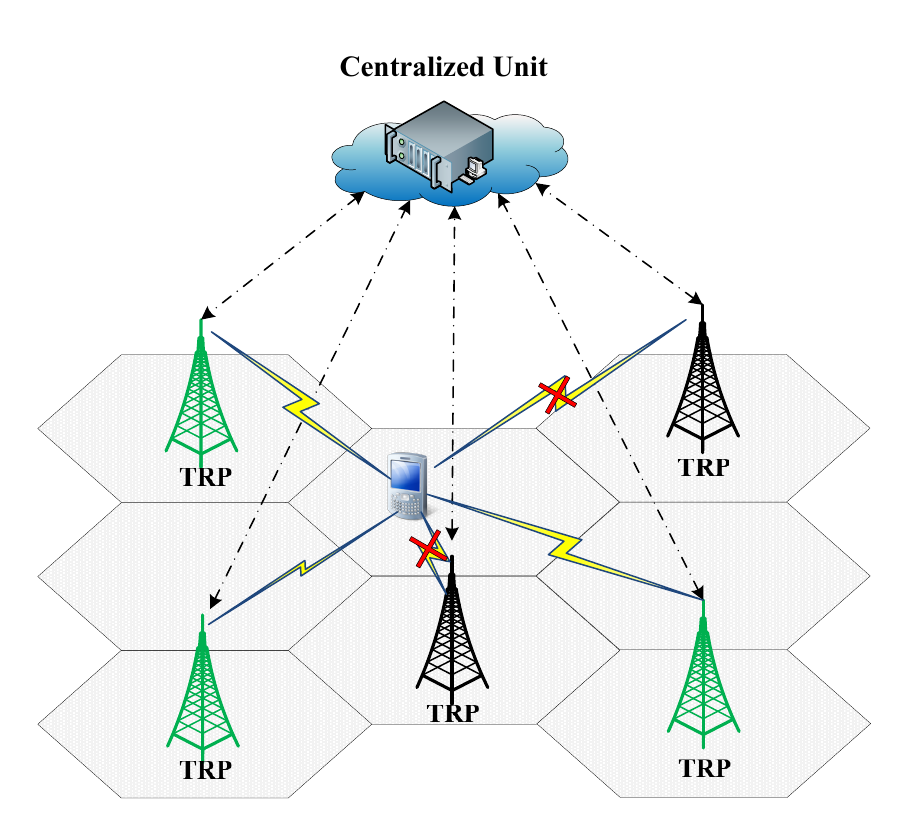}
	\caption{\textcolor{red}{Schematic diagram of the DMIMO system.}}
	\label{fig1}
\end{figure}

\section{System Model}

{\color{red}As shown in \figref{fig1}, we consider a downlink DMIMO system with a centralized control unit (CU) connected to $G$ geographically separated transmission reception points (TRPs), where each TRP is equipped with $M$ antennas \cite{app12178586}. These TRPs cooperatively serve a single UE with $N$ receive antennas. At each slot, the CU selects a subset of $J$ TRPs for joint transmission, where $1 \leq J \leq G$. The total number of candidate node subsets is denoted by $D = \binom{G}{J}$. \footnote{In this letter, ``node'' and ``TRP'' are used interchangeably in the context of DMIMO; that is, cooperative node subset selection is equivalent to TRP subset selection.}}

{\color{red}This single-UE DMIMO model isolates the activation-probability
optimization. All $G$ TRPs belong to the coordinated cluster, and inter-cell
interference from uncoordinated neighboring cells is neglected. It is thus a
controlled DMIMO abstraction of coordinated cooperative
transmission.}

{\color{red}The global channel state information (CSI) is $\boldsymbol{H} = [\boldsymbol{H}^{(1)}, \boldsymbol{H}^{(2)}, \ldots, \boldsymbol{H}^{(G)}] \in \mathbb{C}^{N \times GM}$, where $\boldsymbol{H}^{(g)} \in \mathbb{C}^{N \times M}$ is the channel from the $g$-th TRP to the UE. When the $j$-th TRP subset is selected, the corresponding joint channel is $\boldsymbol{H}_j = \boldsymbol{H}\boldsymbol{S}_j \in \mathbb{C}^{N \times JM}$, where $\boldsymbol{S}_j$ is the corresponding selection matrix. The CU is assumed to have access to the perfect global CSI, and} designs a joint precoding matrix $\boldsymbol{F}_j \in \mathbb{C}^{JM \times N_\mathrm{s}}$ based on $\boldsymbol{H}_j$ to serve $N_\mathrm{s}$ data streams. For multi-stream transmission, we employ singular value decomposition (SVD)-based precoding. The SVD of the joint channel matrix is given by $\boldsymbol{H}_j = \boldsymbol{U}_j \boldsymbol{\Lambda}_j \boldsymbol{V}_j^{\text{H}}$, where $\boldsymbol{U}_j \in \mathbb{C}^{N \times N}$ and $\boldsymbol{V}_j \in \mathbb{C}^{JM \times JM}$ are unitary matrices, and $\boldsymbol{\Lambda}_j \in \mathbb{C}^{N \times JM}$ is a diagonal matrix containing the singular values $\lambda_{j,1} \geq \lambda_{j,2} \geq \cdots \geq \lambda_{j,\min(N,JM)} \geq 0$. Let $\boldsymbol{V}_{j,s} \in \mathbb{C}^{JM \times N_\mathrm{s}}$ denote the submatrix formed by the first $N_\mathrm{s}$ columns of $\boldsymbol{V}_j$, corresponding to the $N_\mathrm{s}$ dominant singular values. The precoding matrix is designed as $\boldsymbol{F}_j = \boldsymbol{V}_{j,s} \boldsymbol{\Gamma}_j^{1/2}$, where $\boldsymbol{\Gamma}_j = \operatorname{diag}(\gamma_{j,1}, \gamma_{j,2}, \ldots, \gamma_{j,N_\mathrm{s}})$ is the power allocation matrix for the $N_\mathrm{s}$ data streams.

\begin{figure}[!t]
       \centering
       \includegraphics[width=0.92\linewidth]{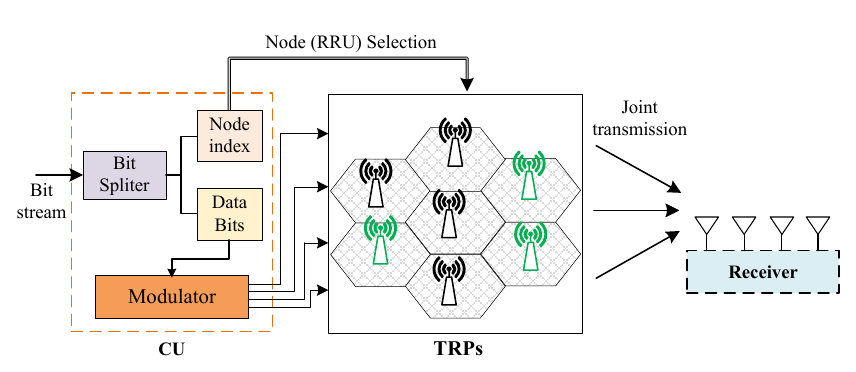}
\caption{\textcolor{red}{System model of non-uniform activation aided DGSM for DMIMO.}}
       \label{Model}
\end{figure}

The received signal is given by
\begin{align}
\boldsymbol{y} = \boldsymbol{H}_j \boldsymbol{F}_j \boldsymbol{x} + \boldsymbol{n},
\end{align}
where $\boldsymbol{x} \in \mathbb{C}^{N_\mathrm{s} \times 1}$ is the vector of transmitted data symbols with $\mathbb{E}\{\boldsymbol{x}\boldsymbol{x}^{\text{H}}\} = \boldsymbol{I}_{N_\mathrm{s}}$, and $\boldsymbol{n} \in \mathbb{C}^{N \times 1}$ is the additive white Gaussian noise (AWGN) vector with $\boldsymbol{n} \sim \mathcal{CN}(\boldsymbol{0}, \sigma_n^2 \boldsymbol{I}_N)$.

In conventional DPS schemes, the optimal node subset is selected based on instantaneous channel state information to maximize the data rate, which can be expressed as
\begin{align}
j^\star = \arg\max_{j \in \{1, 2, \ldots, D\}} R_j,
\end{align}
where $R_j$ represents the achievable data rate when the $j$-th node subset is selected. However, this approach fails to exploit the information-carrying potential of the node selection process itself.

In existing DGSM schemes with uniform activation, each node subset is selected with equal probability, i.e.,
\begin{align}
\Pr(\boldsymbol{S}_j) = \frac{1}{D}, \quad \forall j \in \left\{1, 2, \ldots, D\right\}.
\end{align}
While this maximizes the selection entropy, it ignores the heterogeneous channel quality across various node subsets. Motivated by this limitation, we propose a non-uniform activation probability aided DGSM framework, as illustrated in \figref{Model}, where each node subset $\boldsymbol{S}_j$ is assigned an activation probability $p_j$ based on the channel quality rather than being equally likely. The probability distribution $\boldsymbol{p} = [p_1, p_2, \ldots, p_D]^{\text{T}}$ satisfies
\begin{align}
\sum_{j=1}^{D} p_j = 1, \quad \text{and} \quad p_j \geq 0, \quad \forall j.
\end{align}

The received signal follows a Gaussian mixture distribution conditioned on the node selection, which can be expressed as
\begin{align}
\boldsymbol{y} \sim \sum_{j=1}^{D} p_j \cdot \mathcal{CN}(\boldsymbol{0}, \boldsymbol{\Sigma}_j),
\end{align}
where $\boldsymbol{\Sigma}_j = \boldsymbol{H}_j \boldsymbol{F}_j \boldsymbol{F}_j^{\text{H}} \boldsymbol{H}_j^{\text{H}} + \sigma_n^2 \boldsymbol{I}_N$ represents the covariance matrix of the received signal when the $j$-th TRP subset is selected for transmission.

The fundamental trade-off lies in balancing the information carried by the node selection process (quantified by the selection entropy) and the reliability of data symbol transmission (determined by the channel quality of the selected nodes).

The total mutual information $\mathcal{I}$ between the input and output can be decomposed into two parts\cite{10129110,10804134}:
\begin{align}
\mathcal{I} = \mathcal{I}_{\text{data}} + \mathcal{I}_{\text{selection}},
\end{align}
where $\mathcal{I}_{\text{data}}$ represents the information carried by the data symbols and $\mathcal{I}_{\text{selection}}$ denotes the information carried by the node selection process.

Applying the chain rule of mutual information, we decompose the total mutual information as:
\begin{align}
\mathcal{I} = \mathcal{I}(\boldsymbol{y}; \boldsymbol{x}, j) = \mathcal{I}(\boldsymbol{y}; \boldsymbol{x}|j) + \mathcal{I}(\boldsymbol{y}; j),
\end{align}
where $\mathcal{I}(\boldsymbol{y}; \boldsymbol{x}|j)$ denotes the conditional mutual information given the selected node subset, and $\mathcal{I}(\boldsymbol{y}; j)$ represents the information conveyed by the node selection process.

For MIMO systems with SVD precoding, the conditional mutual information given node selection $j$ is:
\begin{align}
\mathcal{I}(\boldsymbol{y}; \boldsymbol{x}|j) = \sum_{i=1}^{N_\mathrm{s}} \log_2\left(1 + \frac{\gamma_{j,i} \lambda_{j,i}^2}{\sigma_n^2}\right),
\end{align}
where $\lambda_{j,i}$ are the singular values of the channel matrix $\boldsymbol{H}_j$, and $\gamma_{j,i}$ denotes the power allocated to each data stream.

An upper bound for the total mutual information can be derived as \cite{Huber2008OnEA}:
\begin{align}
\mathcal{I}_{\text{upper}} = \sum_{j=1}^{D} p_j \sum_{i=1}^{N_\mathrm{s}} \log_2\left(1 + \frac{\gamma_{j,i} \lambda_{j,i}^2}{\sigma_n^2}\right) + \mathcal{H}(\boldsymbol{p}),
\end{align}
where $\mathcal{H}(\boldsymbol{p}) = -\sum_{j=1}^{D} p_j \log_2 p_j$ is the selection entropy. This upper bound is derived under the assumption of Gaussian signaling.

Based on the above analysis, We develop a comprehensive joint optimization framework encompassing both node selection probabilities and per-stream power allocation across all node subsets.

\section{Problem Formulation and Proposed Solution}
\subsection{Problem Formulation}
\label{subsec:problem_formulation}

The joint optimization problem for maximizing the overall spectral efficiency is formulated as:
\begin{equation}
	\begin{split}
		\max_{\boldsymbol{p}, \boldsymbol{\Gamma}} \quad & \sum_{j=1}^{D} p_j \sum_{i=1}^{N_\mathrm{s}} \log_2\left(1 + \frac{\gamma_{j,i} \lambda_{j,i}^2}{\sigma_n^2}\right) + \mathcal{H}(\boldsymbol{p}) \\
		\text{s.t.} \quad & \sum_{j=1}^{D} p_j = 1, \quad p_j \geq 0, \quad \forall j, \\
		& \sum_{j=1}^{D} p_j \sum_{i=1}^{N_\mathrm{s}} \gamma_{j,i} \leq P_{\text{total}}, \quad \gamma_{j,i} \geq 0, \quad \forall j,i.
	\end{split}
	\label{eq:optimization}
\end{equation}

Problem (\ref{eq:optimization}) maximizes the overall spectral efficiency. The objective function comprises the expected channel capacity and the selection entropy, where the latter promotes diversity in node selection, The optimization is subject to the probability normalization constraint and the long-term average power constraint.

\subsection{SC-JPPO}
\label{subsec:optimization_solution}

We solve the optimization problem using the method of Lagrange multipliers. The Lagrangian function can be expressed as (\ref{lagre}), where $\xi$ and $\mu$ are the Lagrange multipliers associated with the probability normalization constraint $\sum_{j=1}^{D} p_j = 1$ and the long-term average power constraint $\sum_{j=1}^{D} p_j \sum_{i=1}^{N_\mathrm{s}} \gamma_{j,i} \leq P_{\text{total}}$, respectively.

We solve the optimization problem in two steps:

\subsubsection{Step 1: Per-Stream Power Allocation}

Taking the partial derivative of $\mathcal{L}$ with respect to $\gamma_{j,i}$:
\begin{align}
\frac{\partial \mathcal{L}}{\partial \gamma_{j,i}} = \frac{p_j}{\ln 2} \cdot \frac{\lambda_{j,i}^2 / \sigma_n^2}{1 + \gamma_{j,i} \lambda_{j,i}^2 / \sigma_n^2} - \mu p_j = 0.
\end{align}

\begin{figure*}
\begin{align}\label{lagre}
\mathcal{L}(\boldsymbol{p}, \boldsymbol{\Gamma}, \xi, \mu) = \sum_{j=1}^{D} p_j \sum_{i=1}^{N_\mathrm{s}} \log_2\left(1 + \frac{\gamma_{j,i} \lambda_{j,i}^2}{\sigma_n^2}\right) - \sum_{j=1}^{D} p_j \log_2 p_j + \xi\left(1 - \sum_{j=1}^{D} p_j\right) + \mu\left(P_{\text{total}} - \sum_{j=1}^{D} p_j \sum_{i=1}^{N_\mathrm{s}} \gamma_{j,i}\right).
\end{align}
\hrulefill
\vspace*{4pt}
\end{figure*}

For any subset with $p_j > 0$, dividing by $p_j$ yields the well-known water-filling (WF) solution:
\begin{align}
\gamma_{j,i}^\star(\mu) = \left[\frac{1}{\mu \ln 2} - \frac{\sigma_n^2}{\lambda_{j,i}^2}\right]^+,
	\label{eq:waterfilling}
\end{align}
where $[x]^+ = \max(0, x)$ ensures non-negative power allocation. The unified Lagrange multiplier $\mu$ guarantees that all node subsets share a common water level $1/(\mu \ln 2)$.

We denote the per-subset total power and achievable rate as $P_j^\star(\mu) = \sum_{i=1}^{N_\mathrm{s}} \gamma_{j,i}^\star(\mu)$ and $R_j^\star(\mu) = \sum_{i=1}^{N_\mathrm{s}} \log_2(1 + \gamma_{j,i}^\star(\mu) \lambda_{j,i}^2 / \sigma_n^2)$, respectively.

\subsubsection{Step 2: Activation Probability Optimization}

We then optimize the activation probabilities $p_j$. Setting the partial derivative with respect to $p_j$ to zero yields
\begin{align}
\frac{\partial \mathcal{L}}{\partial p_j} = R_j^\star(\mu) - \log_2 p_j - \frac{1}{\ln 2} - \xi - \mu P_j^\star(\mu) = 0,
\end{align}
where the term $-\log_2 p_j - 1/\ln 2$ follows from $\frac{\partial}{\partial p_j}(-p_j \log_2 p_j)$.

Solving for $p_j$ yields the optimal activation probability in closed form:
\begin{equation}\label{eq:activation_prob}
	\begin{split}
		p_j^\star(\mu)
		&=
		\frac{
			2^{R_j^\star(\mu)-\mu P_j^\star(\mu)}
		}{
			\sum_{k=1}^{D}
			2^{R_k^\star(\mu)-\mu P_k^\star(\mu)}
		}
		\\
		&=
		\frac{
			\exp\left(
			\ln2\left[R_j^\star(\mu)-\mu P_j^\star(\mu)\right]
			\right)
		}{
			\sum_{k=1}^{D}
			\exp\left(
			\ln2\left[R_k^\star(\mu)-\mu P_k^\star(\mu)\right]
			\right)
			} \\
			&=
			\mathrm{Softmax}\big(\ln2\big[R_j^\star(\mu)-\mu P_j^\star(\mu)\big]\big).
	\end{split}
\end{equation} The Lagrange multiplier $\mu$ is then determined by solving the one-dimensional equation:
\begin{align}
f(\mu) \triangleq \sum_{j=1}^{D} p_j^\star(\mu) P_j^\star(\mu) - P_{\text{total}} = 0,
	\label{eq:mu_equation}
\end{align}
which depends only on $\mu$. The monotonicity of $f(\mu)$ is established through derivative analysis in Appendix~A, guaranteeing that this equation can be solved efficiently via the bisection method. Note that (\ref{eq:activation_prob}) takes the $\mathrm{Softmax}$ form, which maps the effective utility $\ln2[R_j^\star(\mu)-\mu P_j^\star(\mu)]$ of each subset to a normalized probability.

The overall solution procedure is summarized in \textbf{Algorithm~1}.

\renewcommand{\algorithmicrequire}{\textbf{Input:}}
\renewcommand{\algorithmicensure}{\textbf{Output:}}
\begin{algorithm}[!t]
\caption{SC-JPPO Algorithm}
\begin{algorithmic}[1]
\REQUIRE $\{\boldsymbol{H}_j\}_{j=1}^{D}$,  $\sigma_n^2$, and $P_{\text{total}}$.
\STATE   For each subset $j$, compute $\boldsymbol{H}_j = \boldsymbol{U}_j \boldsymbol{\Lambda}_j \boldsymbol{V}_j^{\text{H}}$ to obtain singular values $\{\lambda_{j,i}\}_{i=1}^{N_\mathrm{s}}$.
\STATE  Find $\mu$ satisfying $f(\mu)=0$ in (\ref{eq:mu_equation}) via the bisection method.
\STATE  Compute $\gamma_{j,i}^\star(\mu)$ for all $j,i$ using (\ref{eq:waterfilling}).
\STATE  Compute $p_j^\star(\mu)$ for all $j$ using (\ref{eq:activation_prob}).
\ENSURE $\{\gamma_{j,i}^\star\}$, $\{p_j^\star\}$.
\end{algorithmic}
\end{algorithm}

{\color{red}
\subsection{Implementation and Complexity Discussion}
\label{subsec:complexity}

The CU centrally performs the optimization once per channel-coherence
interval. For $D=\binom{G}{J}$ candidate subsets, the SVD preprocessing and
$B$ bisection iterations require $\mathcal{O}\!\left(DN(JM)^2+BDN_{\mathrm{s}}\right)$ operations.
In each slot, the CU only sends the selected subset and control commands, and exactly $J$ TRPs are activated.
}

\section{Simulation Results and Discussions}
\label{sec:simulation}

{\color{red}In this section, we evaluate the performance of the proposed non-uniform activation aided DGSM scheme through simulations. The signal-plus-noise ratio (SNR) is denoted as $\frac{P_\text{total}}{\sigma_n^2}$. We consider a DMIMO system with $G = 8$ TRPs, each equipped with $M = 2$ antennas, serving a single user with $N = 8$ receive antennas. Both the transmitter and receiver adopt uniform linear arrays. We select $J = 2$ nodes for cooperative transmission. The channel is modeled as flat Rayleigh fading{\footnote{\color{red}The proposed framework does not rely on a particular channel model. It only requires the corresponding CSI to be available at the CU. In practical DMIMO deployments, each TRP can estimate its local CSI from uplink pilots under time-division duplexing (TDD) reciprocity, and then forward the local CSI to the CU over the fronthaul for global CSI aggregation.}}. The following benchmark schemes are considered:}

\begin{itemize}
    \item \textbf{Best node selection with WF (BNS-WF):} selects the optimal node subset and allocates power via WF;
    \item \textbf{Best node selection with equal power (BNS-EQ):} selects the optimal node subset with equal power allocation;
    \item \textbf{Uniform activation with WF  (UA-WF):} conventional DGSM with uniform activation probabilities and WF per subset.
\end{itemize}

\begin{figure*}[!t]
\centering
\begin{minipage}[t]{0.32\textwidth}
    \centering
    \includegraphics[width=\linewidth]{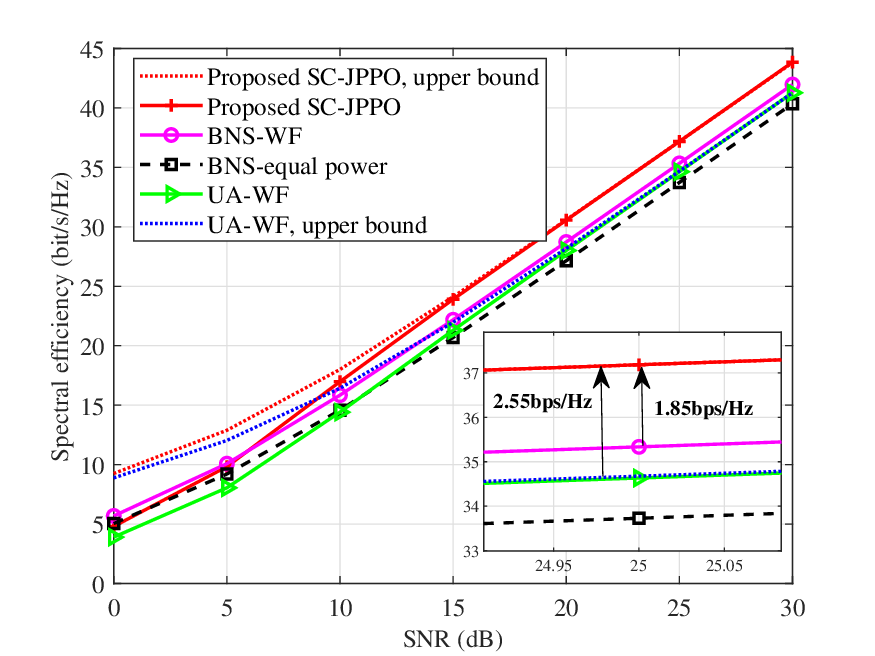}
    \caption{Spectral efficiency comparison versus SNR.}
    \label{SE_comparison}
\end{minipage}\hfill
\begin{minipage}[t]{0.32\textwidth}
    \centering
    \includegraphics[width=\linewidth]{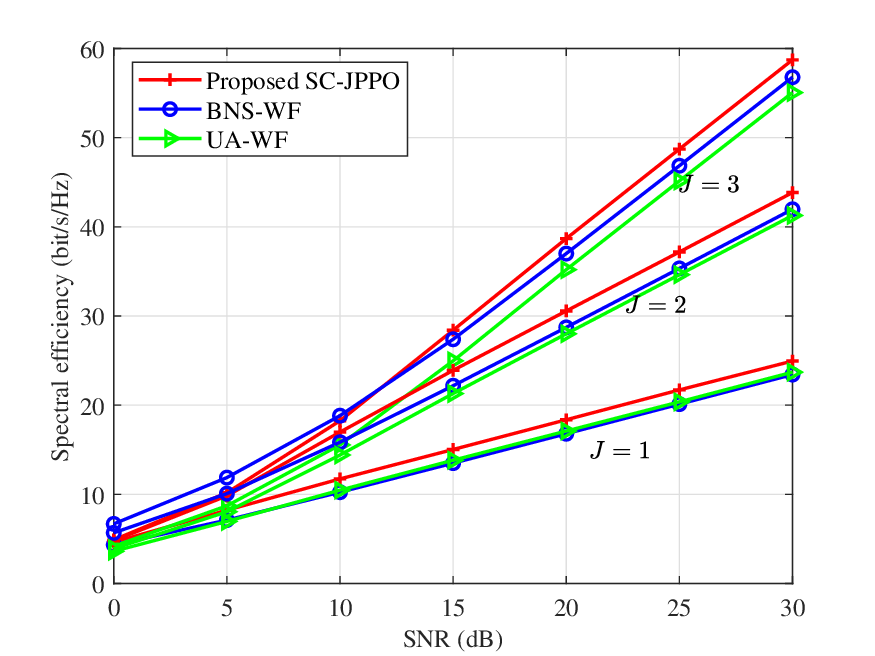}
    \caption{Spectral efficiency comparison versus SNR for different numbers of cooperative nodes $J$.}
    \label{SE_comparison2}
\end{minipage}\hfill
\begin{minipage}[t]{0.32\textwidth}
    \centering
    \includegraphics[width=\linewidth]{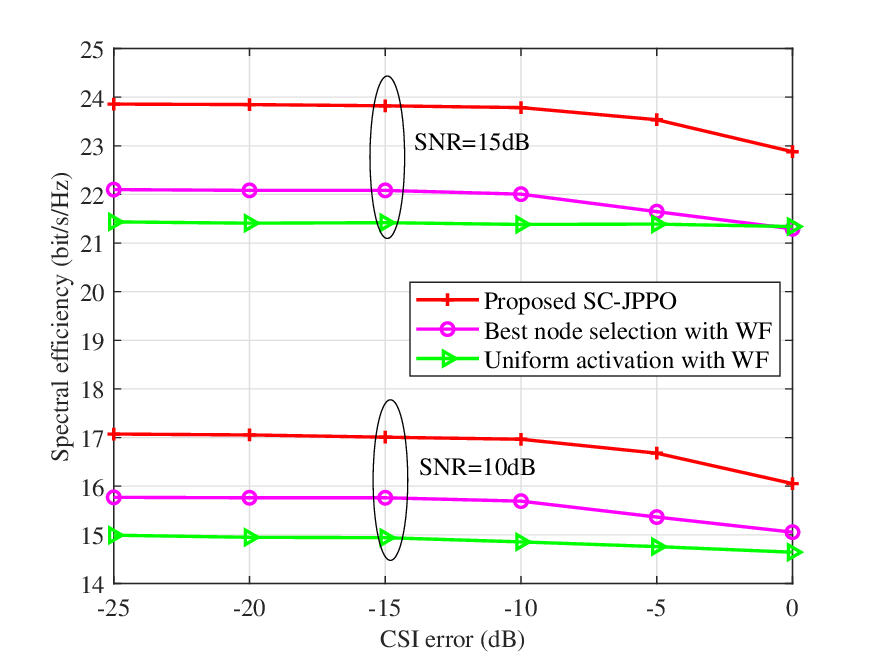}
    \caption{\textcolor{red}{Spectral efficiency comparison versus CSI error level.}}
    \label{SE_comparison3}
\end{minipage}
\end{figure*}

\figref{SE_comparison} compares the spectral efficiency of different schemes versus SNR. As the SNR increases, both the proposed scheme and the uniform activation scheme approach the upper bound, verifying the tightness of the derived bound in the high-SNR regime. The proposed scheme outperforms all benchmarks. At SNR = 25 dB, it achieves a gain of approximately 1.85 bps/Hz over BNS-WF and 2.55 bps/Hz over UA-WF. This improvement stems from the joint optimization of activation probabilities and power allocation, which effectively balances the data rate and the index information rate.

\figref{SE_comparison2} examines the impact of the number of cooperative nodes $J$. In the high-SNR regime, the proposed scheme consistently outperforms the benchmarks, and the gain becomes more significant as $J$ increases. However, in the low-SNR regime, the proposed scheme may exhibit slightly lower performance than BNS-WF in certain cases. This is because the derived upper bound is tight in the high-SNR regime, and the optimization based on this bound may not guarantee optimality under low-SNR conditions.

{\color{red}
To evaluate imperfect CSI, we model the CSI available at the CU as
$\widehat{\boldsymbol{H}}=\boldsymbol{H}+\boldsymbol{H}_{\mathrm{e}}$, where
$\boldsymbol{H}$ is the actual global channel and $\boldsymbol{H}_{\mathrm{e}}$
is an independent estimation error. The error level is measured
by $\mathrm{NMSE}_{\mathrm{dB}}=10\log_{10}(\|\boldsymbol{H}_{\mathrm{e}}\|_{\mathrm{F}}^2/\|\boldsymbol{H}\|_{\mathrm{F}}^2)$. All transmission parameters are designed from
$\widehat{\boldsymbol{H}}$, while the spectral efficiency is evaluated over the
actual channel $\boldsymbol{H}$.
\figref{SE_comparison3} shows the spectral efficiency at a fixed SNR under
varying channel-estimation error levels. As the error level increases, all
schemes experience performance degradation because of subset-selection and
precoding mismatch. Nevertheless, the proposed scheme maintains its advantage
over the considered benchmarks throughout the evaluated error range,
indicating its robustness to imperfect CSI.}

{\color{red}
\begin{remark}
At low SNR, the optimized activation probabilities tend to favor the TRP subset with the highest effective utility, thereby approaching best-node selection and reducing the use of channel-poor subsets relative to uniform activation. This adaptive selection can improve the reliability of the active-set decision. More generally, the spectral-efficiency gain can be exchanged for a lower-order constellation at a fixed target rate, which may improve symbol-detection reliability. A rigorous BER-oriented evaluation requires finite-alphabet mutual-information analysis and joint optimization of the activation probabilities, power allocation, and precoding matrix. These extensions are left for future work.
\end{remark}
}

\section{Conclusion}
\label{sec:conclusion}

{\color{red}This letter proposed a non-uniform activation aided DGSM scheme for a DMIMO
system. A mutual-information upper bound was derived, and a joint activation
probability and power allocation problem was formulated under a long-term
average power constraint. To solve the problem, we developed the SC-JPPO
algorithm, which provides the optimal solution of the upper-bound
maximization problem. The optimal activation probabilities are shown to take
the form of a Softmax function of the post-water-filling effective utilities.
Simulation results confirmed that the proposed scheme improves SE over
uniform activation and conventional best-node selection baselines. {\color{red}
Future work will investigate finite-alphabet inputs and multi-user distributed
transmission.}}

\appendices
\section{Monotonicity of $f(\mu)$}
\label{app:monotonicity}


Let $\mathcal A_j(\mu)=\{i:\gamma_{j,i}^\star(\mu)>0\}$ denote the active
stream set of subset $j$. For $i\in\mathcal A_j(\mu)$, we have
$\gamma_{j,i}^\star(\mu)=\frac{1}{\mu\ln2}-\frac{\sigma_n^2}{\lambda_{j,i}^2}$, and hence
$\frac{{\rm d}\gamma_{j,i}^\star(\mu)}{{\rm d}\mu}=-\frac{1}{\mu^2\ln2}$. Therefore,
$P_j^\star(\mu)=\sum_{i\in\mathcal A_j(\mu)}\gamma_{j,i}^\star(\mu)$ and
$$
	\frac{{\rm d}P_j^\star(\mu)}{{\rm d}\mu}
	=
	-\frac{|\mathcal A_j(\mu)|}{\mu^2\ln2}.$$

For the active streams, the achievable rate of subset $j$ can be written as
$R_j^\star(\mu)=\sum_{i\in\mathcal A_j(\mu)}
\log_2\!\Big(\frac{\lambda_{j,i}^2}{\mu\sigma_n^2\ln2}\Big)$, which gives
\begin{align*}
	\frac{{\rm d}R_j^\star(\mu)}{{\rm d}\mu}
	=
	-\frac{|\mathcal A_j(\mu)|}{\mu\ln2}.
\end{align*}

Define
$\phi_j(\mu)=\big[R_j^\star(\mu)-\mu P_j^\star(\mu)\big]\ln2$.
Then the optimal activation probability can be expressed as
$p_j^\star(\mu)=\dfrac{\exp(\phi_j(\mu))}{\sum_{k=1}^{D}\exp(\phi_k(\mu))}$. Then, we have
\begin{align*}
	\frac{{\rm d}\phi_j(\mu)}{{\rm d}\mu}
	&=
	\left[
	\frac{{\rm d}R_j^\star(\mu)}{{\rm d}\mu}
	-
	P_j^\star(\mu)
	-
	\mu
	\frac{{\rm d}P_j^\star(\mu)}{{\rm d}\mu}
	\right]\ln2
	=
	-P_j^\star(\mu)\ln2 .
\end{align*}
Thus, using the derivative of the softmax function, we obtain
\begin{align*}
	\frac{{\rm d}p_j^\star(\mu)}{{\rm d}\mu}
	&=
	p_j^\star(\mu)
	\left[
	\frac{{\rm d}\phi_j(\mu)}{{\rm d}\mu}
	-
	\sum_{k=1}^{D}p_k^\star(\mu)
	\frac{{\rm d}\phi_k(\mu)}{{\rm d}\mu}
	\right]  \\
	&=
	p_j^\star(\mu)\ln2
	\left[
	\bar P(\mu)-P_j^\star(\mu)
	\right],
\end{align*}
where
$\bar P(\mu)=\sum_{k=1}^{D}p_k^\star(\mu)P_k^\star(\mu)$.

Finally, differentiating $f(\mu)$ gives
\begin{align*}
	\frac{{\rm d}f(\mu)}{{\rm d}\mu}
	&=
	\sum_{j=1}^{D}
	\left[
	\frac{{\rm d}p_j^\star(\mu)}{{\rm d}\mu}P_j^\star(\mu)
	+
	p_j^\star(\mu)
	\frac{{\rm d}P_j^\star(\mu)}{{\rm d}\mu}
	\right]\\
	&=
	\ln2
	\left[
	\bar P^2(\mu)
	-
	\sum_{j=1}^{D}
	p_j^\star(\mu)
	\left(P_j^\star(\mu)\right)^2
	\right] \\
	&\quad
	-
	\frac{1}{\mu^2\ln2}
	\sum_{j=1}^{D}
	p_j^\star(\mu)|\mathcal A_j(\mu)|.
\end{align*}
Since
$\sum_{j=1}^{D}p_j^\star(\mu)(P_j^\star(\mu))^2-\bar P^2(\mu)\geq0$ and the second term above
is strictly negative
we have ${\rm d}f(\mu)/{\rm d}\mu\leq0$. Therefore, $f(\mu)$ is monotonically non-increasing
with respect to $\mu$. 

\bibliographystyle{IEEEtran}

\bibliography{IEEEabrv,bib}

\begin{thebibliography}{10}
\providecommand{\url}[1]{#1}
\csname url@samestyle\endcsname
\providecommand{\newblock}{\relax}
\providecommand{\bibinfo}[2]{#2}
\providecommand{\BIBentrySTDinterwordspacing}{\spaceskip=0pt\relax}
\providecommand{\BIBentryALTinterwordstretchfactor}{4}
\providecommand{\BIBentryALTinterwordspacing}{\spaceskip=\fontdimen2\font plus
\BIBentryALTinterwordstretchfactor\fontdimen3\font minus \fontdimen4\font\relax}
\providecommand{\BIBforeignlanguage}[2]{{%
\expandafter\ifx\csname l@#1\endcsname\relax
\typeout{** WARNING: IEEEtran.bst: No hyphenation pattern has been}%
\typeout{** loaded for the language `#1'. Using the pattern for}%
\typeout{** the default language instead.}%
\else
\language=\csname l@#1\endcsname
\fi
#2}}
\providecommand{\BIBdecl}{\relax}
\BIBdecl

\bibitem{Lee2012CoMPDeploy}
D.~Lee, H.~Seo, B.~Clerckx, E.~Hardouin, D.~Mazzarese, S.~Nagata, and K.~Sayana, ``Coordinated multipoint transmission and reception in {LTE-Advanced}: Deployment scenarios and operational challenges,'' \emph{IEEE Commun. Mag.}, vol.~50, no.~2, pp. 148--155, Feb. 2012.

\bibitem{Lee2012CoMPSystems}
J.~Lee, Y.~Kim, H.~Lee, B.~L. Ng, D.~Mazzarese, J.~Liu, W.~Xiao, and Y.~Zhou, ``Coordinated multipoint transmission and reception in {LTE-Advanced} systems,'' \emph{IEEE Commun. Mag.}, vol.~50, no.~11, pp. 44--50, Nov. 2012.

\bibitem{Sun2013CoMP}
S.~Sun, Q.~Gao, Y.~Peng, Y.~Wang, and L.~Song, ``Interference management through {CoMP} in {3GPP LTE-Advanced} networks,'' \emph{IEEE Wireless Commun.}, vol.~20, no.~1, pp. 59--66, Feb. 2013.

\bibitem{Mesleh2008SM}
R.~Mesleh, H.~Haas, S.~Sinanovi{\'c}, C.~W. Ahn, and S.~Yun, ``Spatial modulation,'' \emph{IEEE Trans. Veh. Technol.}, vol.~57, no.~4, pp. 2228--2241, Jul. 2008.

\bibitem{DiRenzo2011SMSurvey}
M.~D. Renzo, H.~Haas, and P.~M. Grant, ``Spatial modulation for multiple-antenna wireless systems: A survey,'' \emph{IEEE Commun. Mag.}, vol.~49, no.~12, pp. 182--191, Dec. 2011.

\bibitem{Wen2019SMSurvey}
M.~Wen, B.~Zheng, K.~J. Kim, M.~D. Renzo, T.~A. Tsiftsis, K.-C. Chen, and N.~Al-Dhahir, ``A survey on spatial modulation in emerging wireless systems: Research progresses and applications,'' \emph{IEEE J. Sel. Areas Commun.}, vol.~37, no.~9, pp. 1949--1972, Sep. 2019.

\bibitem{Younis2010GSM}
A.~Younis, N.~Serafimovski, R.~Mesleh, and H.~Haas, ``Generalised spatial modulation,'' in \emph{Proc. Asilomar Conf. Signals, Syst., Comput.}, Nov. 2010, pp. 1498--1502.

\bibitem{Wang2012MASM}
J.~Wang, S.~Jia, and J.~Song, ``Generalised spatial modulation system with multiple active transmit antennas and low complexity detection scheme,'' \emph{IEEE Trans. Wireless Commun.}, vol.~11, no.~4, pp. 1605--1615, Apr. 2012.

\bibitem{Narasimhan2015GSM}
T.~L. Narasimhan, P.~Raviteja, and A.~Chockalingam, ``Generalized spatial modulation in large-scale multiuser {MIMO} systems,'' \emph{IEEE Trans. Wireless Commun.}, vol.~14, no.~7, pp. 3764--3779, Jul. 2015.

\bibitem{10978803}
C.~Zhang, H.~Xu, B.~K. Ng, C.-T. Lam, and K.~Wang, ``Ris-assisted received adaptive spatial modulation for wireless communications,'' in \emph{2025 IEEE Wireless Communications and Networking Conference (WCNC)}, 2025, pp. 1--6.

\bibitem{11506249}
C.~Zhang, B.~K. Ng, K.~Wang, H.~Xu, and C.-T. Lam, ``From reliability to security: How ris-assisted adaptive {SM} and {SSK} enhances wireless systems,'' \emph{IEEE Trans. Veh. Technol. (early access)}, May 2026.

\bibitem{Narayanan2016DSM}
S.~Narayanan, M.~D. Renzo, F.~Graziosi, and H.~Haas, ``Distributed spatial modulation: A cooperative diversity protocol for half-duplex relay-aided wireless networks,'' \emph{IEEE Trans. Veh. Technol.}, vol.~65, no.~5, pp. 2947--2964, May 2016.

\bibitem{Elganimi2020DGSM}
T.~Y. Elganimi, F.~I. Alwerfly, and A.~A. Marseet, ``Distributed generalized spatial modulation for relay networks,'' in \emph{Proc. 28th Signal Process. Commun. Appl. Conf.}, Oct. 2020, pp. 1--4.

\bibitem{app12178586}
\BIBentryALTinterwordspacing
A.~M. Awad, M.~Shehata, S.~M. Gasser, and H.~EL-Badawy, ``Comp-aware bbu placements for 5g radio access networks over optical aggregation networks,'' \emph{Appl. Sci.}, vol.~12, no.~17, Aug. 2022. [Online]. Available: \url{https://www.mdpi.com/2076-3417/12/17/8586}
\BIBentrySTDinterwordspacing

\bibitem{10129110}
S.~Guo and K.~Qu, ``Beamspace modulation for near field capacity improvement in {XL}-{MIMO} communications,'' \emph{{IEEE} Wireless Commun. Lett.}, vol.~12, no.~8, pp. 1434--1438, Aug. 2023.

\bibitem{10804134}
K.~Qu, K.~Pang, H.~Zhao, A.~Elzanaty, and S.~Guo, ``Distance-aware beamspace modulation for near-field spectral efficiency improvement,'' \emph{{IEEE} Wireless Commun. Lett.}, vol.~14, no.~3, pp. 636--640, Aug. 2025.

\bibitem{Huber2008OnEA}
M.~F. Huber, T.~Bailey, H.~F. Durrant-Whyte, and U.~D. Hanebeck, ``On entropy approximation for gaussian mixture random vectors,'' \emph{2008 IEEE International Conference on Multisensor Fusion and Integration for Intelligent Systems}, pp. 181--188, Oct. 2008.

\end{thebibliography}

\end{document}